\documentclass[epj]{svjour}

\usepackage{graphicx}
\usepackage{fancyhdr}
\def\makeheadbox{{
 }}
\def\mytitle{My title}
\def\myauthors{My name}
\def\mytype{My type of session}
\def\mysession{My session}

\def\mytitle{Search for Rare b-Hadron Decays at CDF II} 
\def\myauthors{Philipp Mack}    
\def\mytype{Contributed Talk}
\def\mysession{Flavor Physics}

\begin{document}
\title{Search for rare b-meson decays at CDF}
\author{Philipp Mack (For the CDF Collaboration)
}                     
\institute{Institut f\"ur Experimentelle Kernphysik, University of Karlsruhe, Wolfgang-Gaede-Str. 1, 76131 Karlsruhe, Germany}
\date{}
\abstract{
We report on the search for $B^0_s\rightarrow\mu^+\mu^-$, $B^0_d\rightarrow\mu^+\mu^-$ decays and
$b\rightarrow s \mu^+\mu^-$ transitions in exclusive decays of B mesons using the CDF II detector at the
Fermilab Tevatron Collider. Using 2 fb$^{-1}$ of Run II data we find upper limits on the
branching fractions $\mathcal{B}(B^0_s\rightarrow\mu^+\mu^-)<5.8\times 10^{-8}$ and
$\mathcal{B}(B^0_d\rightarrow\mu^+\mu^-)<1.8\times 10^{-8}$ at 95$\%$ confidence level.
The results for the branching fractions of the $b\rightarrow s \mu^+\mu^-$ transitions using 924 pb$^{-1}$ of Run II data are $\mathcal{B}(B^+\rightarrow \mu^+\mu^-K^+)=(0.60\pm0.15\pm0.04)\times 10^{-6}$, $\mathcal{B}(B^0_d\rightarrow \mu^+\mu^-K^{*0})=(0.82\pm0.31\pm0.10)\times 10^{-6}$ and $\mathcal{B}(B^0_s\rightarrow \mu^+\mu^-\phi)/\mathcal{B}(B^0_s\rightarrow J/\psi\phi)<2.61\times 10^{-3}$ at 95$\%$ confidence level.
} 
\maketitle
\section{Introduction}
\label{intro}
The decay of a $b$ quark into two muons, as well as in an $s$ quark and two muons, requires a
flavor-changing neutral current (FCNC) process which is highly suppressed in the
standard model (SM) as they can only occur through higher order diagrams.
New physics can significantly enhance the branching fractions of these decays. In these proceedings
the current results of the CDF experiment are presented for the branching ratios of
the rare decays  $B^0_{(s,d)}\rightarrow\mu^+\mu^-$ and $B\rightarrow \mu^+\mu^- h$, where
$B$ stands for $B^+$, $B^0_d$, or $B_s$, and h stands for $K^+$,$K^{*0}$ or $\phi$.
The $K^{*0}$ is reconstructed in the mode $K^{*0}\rightarrow K^+\pi^-$ and the
$\phi$ is reconstructed as $\phi\rightarrow K^+K^-$. A detailed description of the
analyses can be found in Ref. \cite{RefJ2,RefJ1}.
\section{The CDF II Detector}
\label{detector}
The CDF II detector is a cylindrical general-purpose particle detector built at one
of the two collision points of the Tevatron $p\bar{p}$ collider which operates at
a center-of-mass energy of $\sqrt{s}=1.96$ TeV. Its inner tracking system consists
of a silicon microstrip detector surrounded by an open-cell wire drift
chamber. The tracking systems are immersed in a 1.4 T magnetic field
and measure the momentum of charged particles.
The electromagnetic and hadronic sampling calorimeters are located outside the solenoid.
The outermost part of the CDF II detector is the muon detector system.
Muons are detected by the planar drift chambers (CMU) 
and the central muon extension (CMX), which consists of conical sections of drift tubes.
The CMU covers a pseudorapidity range up to $|\eta|<0.6$, where
$\eta=-\ln(\tan{\frac{\theta}{2}})$ and $\theta$ is the angle of the track with respect
to the beamline, selecting muons with a $p_T>1.4$ GeV/c.
The CMX extends the coverage to a pseudorapidity range of $0.6<|\eta|<1.0$ for muons
with $p_T>2.0$ GeV/c.
\section{Search Methodology}
The searches for the rare decays $B^0_{(s,d)}\rightarrow\mu^+\mu^-$ and $B\rightarrow \mu^+\mu^- h$
use in both cases a similar approach. A data sample with an integrated luminosity of 2 fb$^{-1}$ is used for $B^0_{(s,d)}\rightarrow\mu^+\mu^-$ search, respectively 924 pb$^{-1}$ for the $B\rightarrow \mu^+\mu^- h$ search.
In both cases events are selected by the dimuon trigger. 
For the $B^0_{(s,d)}\rightarrow\mu^+\mu^-$ analysis the data is futher divided into two classes. Either
both muons are reconstructed in the CMU chambers, further called CMU-CMU, or one muon is
reconstructed in the CMX chambers and the other in the CMU chambers, called CMU-CMX.
\begin{figure}[b]
\begin{center}
\begin{minipage}[b]{0.285\textwidth}
\centering
\includegraphics[width=1\textwidth]{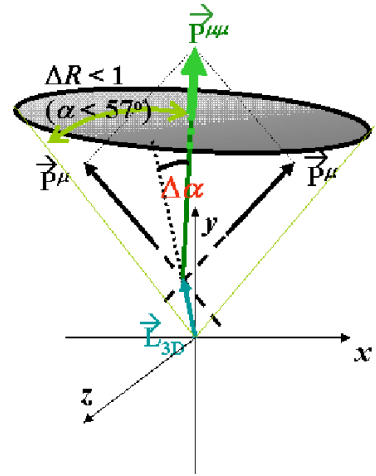}
\end{minipage}
\caption{Graphical representation of the discriminating variables.\label{fig:10}}
\end{center}
\end{figure}
\subsection{Selection Optimization}
To optimize the data selection a signal event sample from MC
simulation and a background sample from data sidebands is used.
\\In case of the $B^0_{(s,d)}\rightarrow\mu^+\mu^-$ analysis a
multivariate neural network (NN) enhances the signal and
background separation. It is based on the discriminating variables
illustrated in Figure \ref{fig:10}: the proper decay length
$\lambda=\beta \gamma c\tau$, the 3D opening angle $\Delta\theta$ between
the dimuon momentum $\vec{p}^{\mu\mu}$ and the displacement
vector from the primary to the dimuon vertex $\vec{L}$, the $p_T$ of the
lower momentum muon candidate and the B-candidate isolation
$I$ \cite{ISO}. 
\\ In case of the
$B\rightarrow \mu^+\mu^- h$ analysis a cut based approach is used
to optimize the selection for the figure of merit $S/\sqrt{S+B}$, 
where $S$ is the estimate of the expected
yield of the rare decays and $B$ is the expected background. The
discriminating variables are the decay length significance $\lambda/\sigma_{\lambda}$, the
pointing angle $\alpha$ from the B meson candidate to the primary
vertex and the isolation $I$. For the final selection candidates with a dimuon mass 
near the $J/\psi$ and the $\psi^{\prime}$ are rejected. 
\subsection{Normalization Modes}
In order to obtain the branching ratios of the different rare
decays, one normalizes to the $B\rightarrow J/\psi h$ modes. The
branching ratios for the decays $B\rightarrow \mu^+\mu^- h$ are
then given by
\begin{eqnarray}
\label{eqn1}
\frac{\mathcal{B}(B\rightarrow\mu^+\mu^-
h)}{\mathcal{B}(B\rightarrow J/\psi h)} & = &\frac{N_{\mu^+\mu^-
h}}{N_{J/\psi h}}\frac{\epsilon_{J/\psi h}}{\epsilon_{\mu^+\mu^-
h}}\cdot \mathcal{B}(J/\psi \rightarrow\mu^+\mu^-)\nonumber \\
\end{eqnarray}
where $N_{\mu^+\mu^- h}$ is the observed number of $B\rightarrow
\mu^+\mu^- h$ decays, $N_{J/\psi h}$ is the observed number of
$B\rightarrow J/\psi h$ decays, while $\epsilon_{J/\psi h}$ and
$\epsilon_{\mu^+\mu^- h}$ are the selection efficiencies of
$B\rightarrow J/\psi h$ and $B\rightarrow \mu^+\mu^- h$
respectively. 
The ratio of efficiencies is about 70 to 85 \% \cite{RefJ2}.\\
In case of the rare decays $B^0_{(s,d)}\rightarrow\mu^+\mu^-$ the
upper limit on the branching fraction can be expressed as
\begin{eqnarray}
\label{eqn2}
\mathcal{B}(B^0_{s,d}\rightarrow\mu^+\mu^-)^{95\%\mbox{\tiny
C.L.}} & = &
\frac{N^{95\%}_{B^0_{s,d}}}{N_{B^+}}\cdot\frac{\alpha_{B^+}}{\alpha_{B^0_{s,d}}}\cdot\frac{\epsilon^{\mbox{\tiny
base}}_{B^+}}{\epsilon^{\mbox{\tiny
base}}_{B^0_{s,d}}}\cdot\frac{1}{\epsilon^{\mbox{\tiny
NN}}_{B^0_{s,d}}}\nonumber \\ & & \cdot \frac{f_u}{f_{s,d}}\cdot
\mathcal{B}(B^+\rightarrow J/\psi K^+)\nonumber \\
\end{eqnarray} 
where $N^{95\%}_{B^0_{s,d}}$ is the upper limit on the number of
$B^0_{s,d}\rightarrow \mu^+\mu^-$ decays at the 95$\%$ C.L. determined from the comparison between expected and 
observed background events,
$N_{B^+}$ is the number of reconstructed $B^+\rightarrow J/\psi
K^+$ candidates. The parameters $\alpha$, $\epsilon^{\mbox{\tiny
base}}$ and $\epsilon^{\mbox{\tiny{NN}}}$ are the trigger acceptances and the efficiencies of
the initial, respectively the NN, requirements.
$\frac{f_u}{f_{s,d}}$ denotes the ratio between the probabilities that a $b$ 
quark produced in a ppbar collisions hadronizes into a $B^+$ meson and a 
$B_s$ or $B_d$ meson \cite{RefJ1}.
\subsection{Background Estimation}
To estimate the background contributions several different sources are considered.\\
In case of the rare decays $B\rightarrow \mu^+\mu^- h$ these sources are
charmless B decays into charged hadrons, reflections between the three rare decay modes
and combinatorial background. The background originating from charmless B decays is calculated 
from a simulation of these decays convoluted with the misidentification rates of muon
detectors obtained from $D^*$-tagged $D^0\rightarrow K^-\pi^+$ decays. 
The combinatorial background is estimated from the high mass sidebands of the $B$ signal
and extrapolated under the signal region using the background shape from the data distribution
with poor vertex quality.\\ In case of the $B^0_{s,d}\rightarrow\mu^+\mu^-$ decays
the background consists of contributions from $B^0_{s,d}\rightarrow h^+ h^-$, where $h^{\pm} =
\pi^{\pm}$ or $K^{\pm}$, and combinatoric background. The expected background from $B^0_{s,d}\rightarrow h^+ h^-$
is calculated from equation \ref{eqn2} by replacing $B^0_{s,d}\rightarrow \mu^+ \mu^-$ with  
$B^0_{s,d}\rightarrow h^+ h^-$ and including two additional efficiency terms to account for the muon misidentification
rates, measured in $D^0\rightarrow K\pi$ data, and the fraction of misidentified $B^0_{s,d}\rightarrow h^+ h^-$
events falling in the signal windows. The branching ratio for the various $B^0_{s,d}\rightarrow h^+ h^-$ 
modes are taken from Ref. \cite{RefJ3}. The combinatoric background is estimated by extrapolating
the number of events in the sideband regions passing a given cut to the signal region using a linear fit.
The total background is then formed by summing up the combinatoric background and the contributions from 
$B^0_{s,d}\rightarrow h^+ h^-$ decays.
\section{Results}
\subsection{$B\rightarrow \mu^+\mu^- h$}
\begin{figure}[hb]
\begin{center}
\begin{minipage} [bth] {0.30\textwidth}
\includegraphics[width=1\textwidth,angle=0]{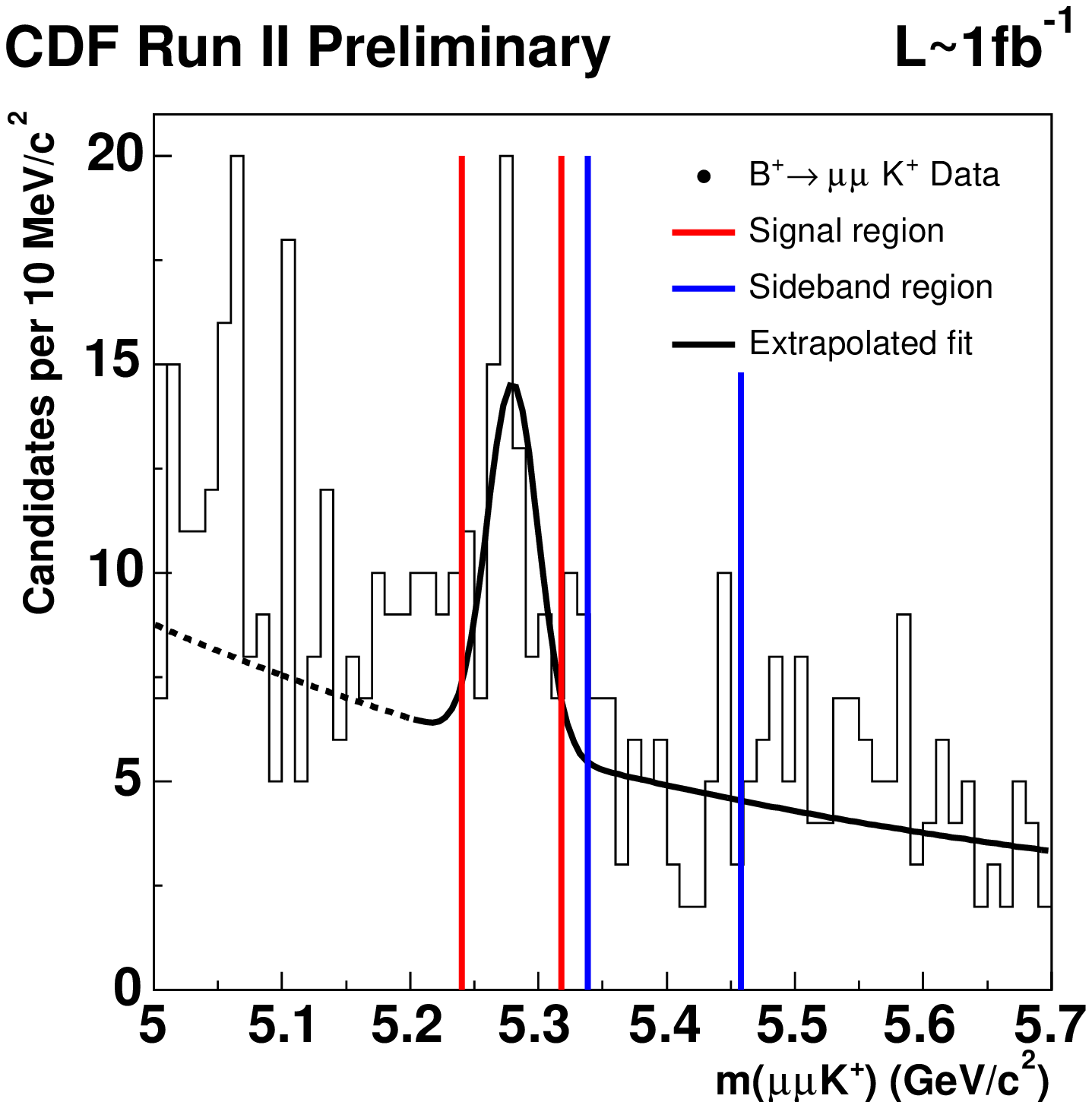}
\end{minipage}
\caption[]{Inv. mass spectrum for the decay $B^+\rightarrow \mu^+\mu^- K^+$.\footnotemark[1] \label{fig:21}}
\begin{minipage} [bth] {0.30\textwidth}
\includegraphics[width=1\textwidth,angle=0]{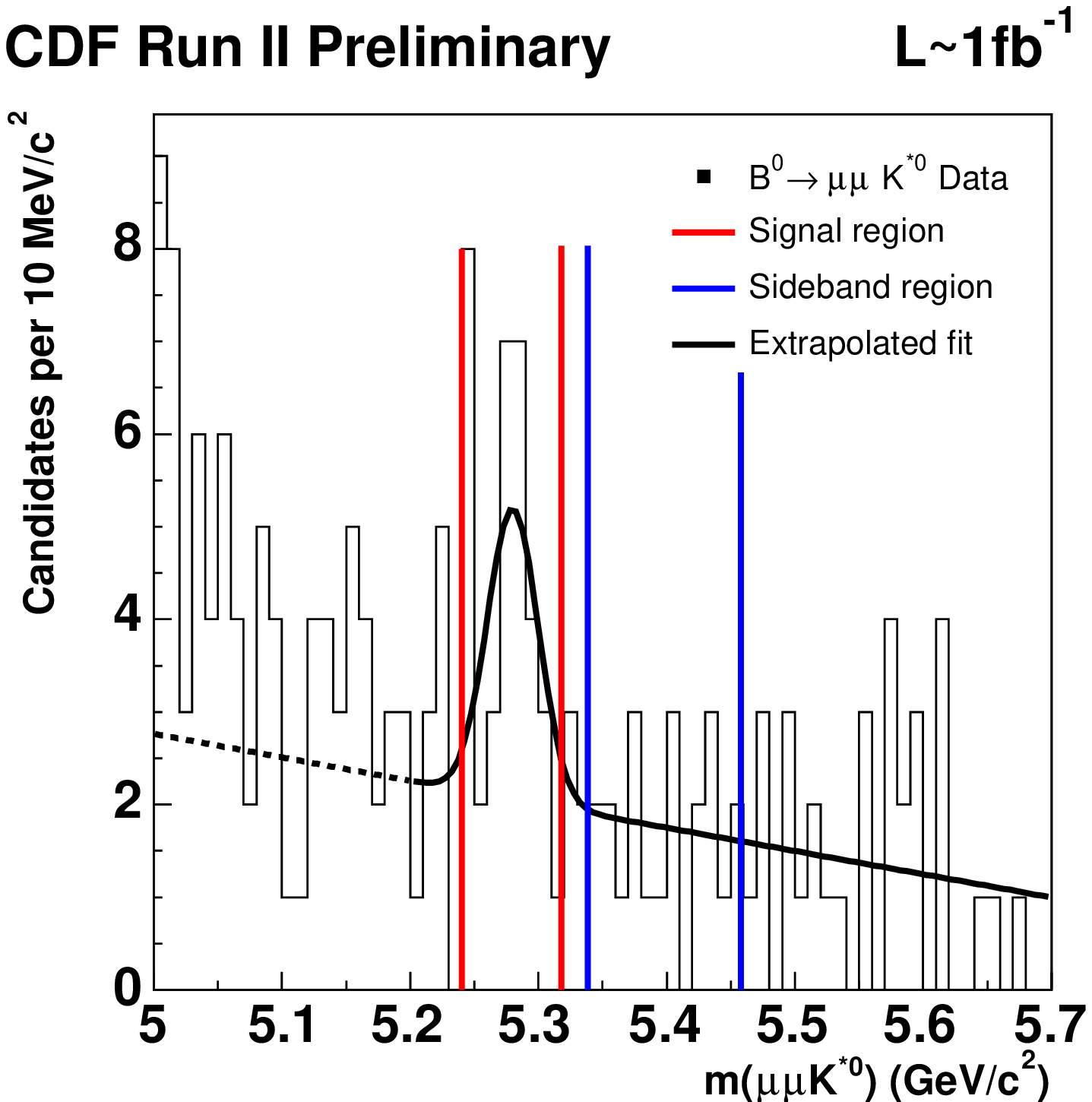}
\end{minipage}
\caption[]{Inv. mass spectrum for the decay $B^0_d\rightarrow \mu^+\mu^- K^{*0}$.\footnotemark[1]\label{fig:22}}
\begin{minipage} [bth] {0.30\textwidth}
\includegraphics[width=1\textwidth,angle=0]{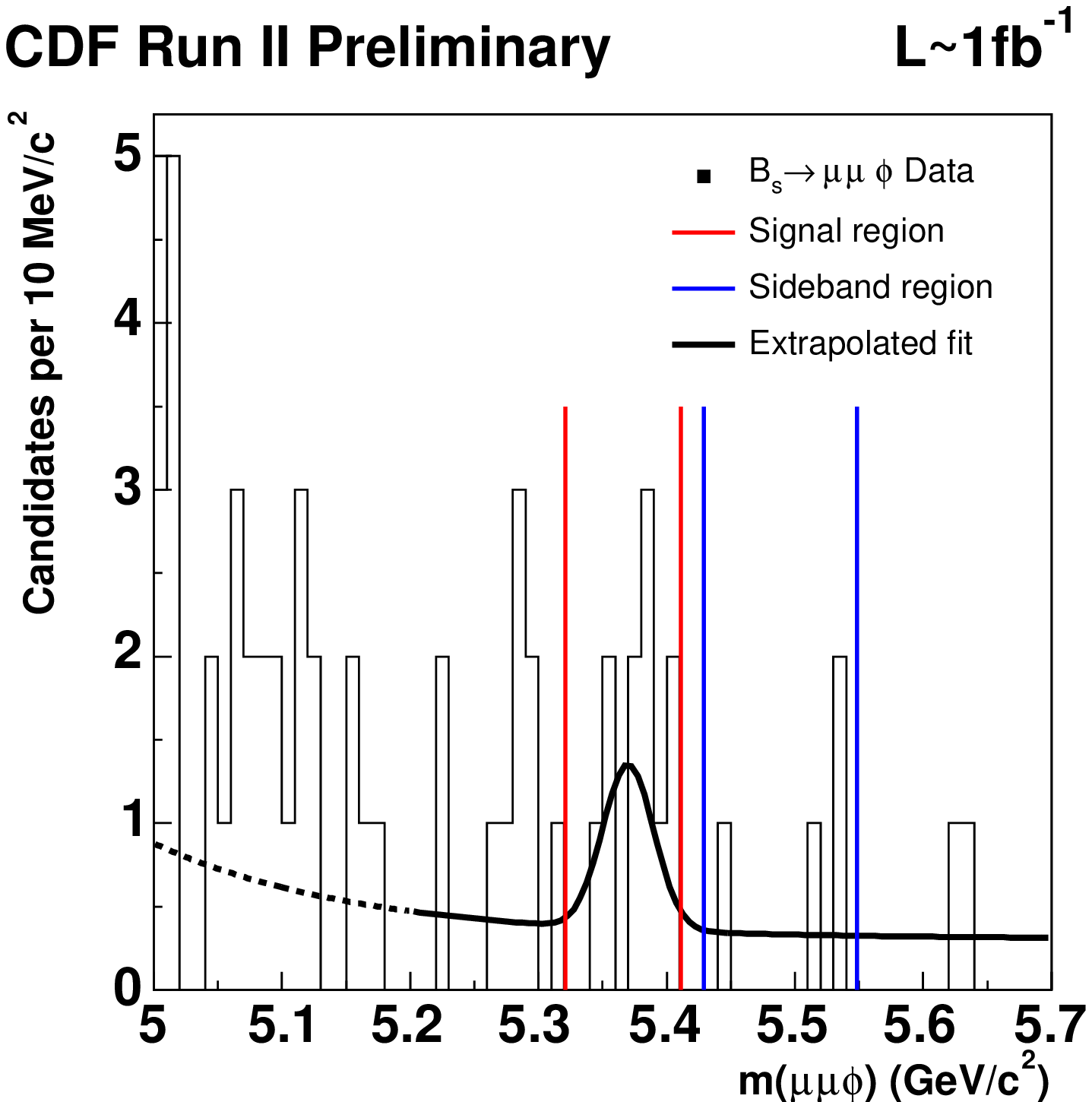}
\end{minipage}
\caption[]{Inv. mass spectrum for the decay $B^0_s\rightarrow \mu^+\mu^- \phi$.\footnotemark[1] \label{fig:23}} 
\end{center}
\end{figure}

In case of the rare decays $B\rightarrow \mu^+\mu^- h$ we observe
an excess in the signal region in all three decay modes. The invariant mass distributions are
shown in Figures \ref{fig:21}, \ref{fig:22} and \ref{fig:23}. The significance of the excess is 
determined by calculating the probability for the background to fluctuate into the number of
observed events. The results for the number of observed events, expected background events, the significance
and the absolute and relative branching ratios are listed in table \ref{tab:2}.
The branching ratios are calculated using equation \ref{eqn1}. We find 
\begin{eqnarray}
\frac{\mathcal{B}(B^+\rightarrow \mu^+\mu^- K^+)}{\mathcal{B}(B^+\rightarrow J/\psi K^+)} & = & (0.59\pm0.15\pm0.03)\times 10^{-3} 
\nonumber \\
\frac{\mathcal{B}(B^0_d\rightarrow \mu^+\mu^- K^{*0})}{\mathcal{B}(B^0_d\rightarrow J/\psi K^{*0})} & = & (0.62\pm0.23\pm0.07)\times 10^{-3} \nonumber \\
\frac{\mathcal{B}(B^0_s\rightarrow \mu^+\mu^- \phi)}{\mathcal{B}(B^0_s\rightarrow J/\psi \phi)}&=&(1.24\pm0.60\pm0.15)\times 10^{-3}
\nonumber 
\end{eqnarray}
Using the world average branching ratios of the normalization modes \cite{Yao}, we find
\begin{eqnarray}
\mathcal{B}(B^+\rightarrow \mu^+\mu^- K^+) & = & (0.60\pm0.15\pm0.04)\times 10^{-6}\nonumber \\
\mathcal{B}(B^0_d\rightarrow \mu^+\mu^- K^{*0}) &= &(0.82\pm0.31\pm0.10)\times 10^{-6}\nonumber \\
\mathcal{B}(B^0_s\rightarrow \mu^+\mu^- \phi) & = & (1.16\pm0.56\pm0.42)\times 10^{-6} \nonumber 
\end{eqnarray}
Since the excess in $B^0_s\rightarrow \mu^+\mu^- \phi$ is not significant, we calculate a limit on its 
relative branching ratio using Bayesian integration assuming a flat prior, and find at $95(90)\%$ C.L.
\begin{eqnarray} 
\frac{\mathcal{B}(B^0_s\rightarrow \mu^+\mu^- \phi)}{\mathcal{B}(B^0_s\rightarrow J/\psi \phi)} &<& 2.61(2.30)\times 10^{-3}
\nonumber
\end{eqnarray} 
\begin{figure}[b]
\begin{center}
\begin{minipage} [bht] {0.42\textwidth}
\includegraphics[width=1\textwidth,angle=0]{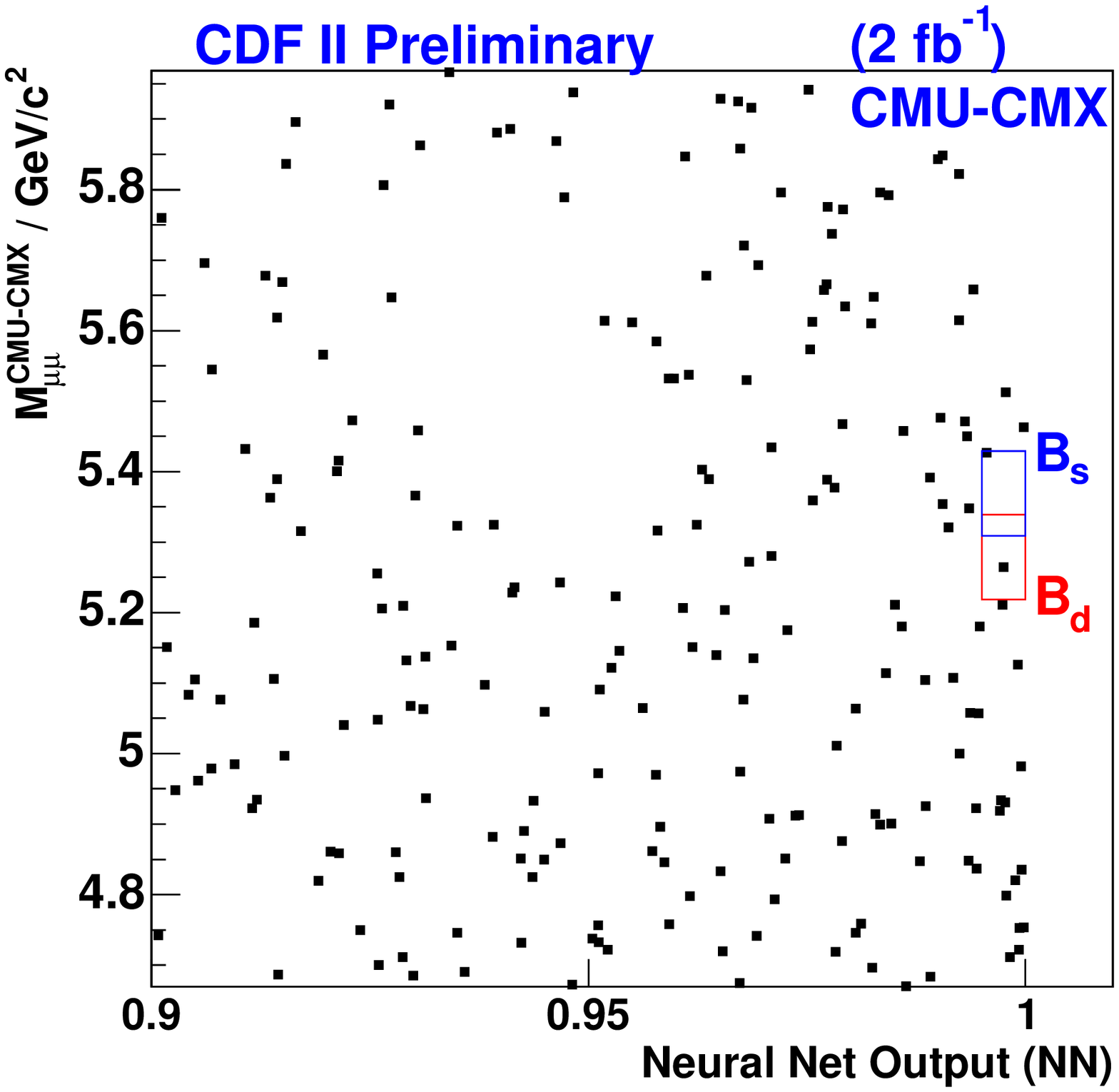}
\end{minipage}
\caption{The invariant mass distribution versus the NN output for the CMU-CMX channel\label{fig:11}.}
\vspace{1.0cm}
\begin{minipage} [bht] {0.42\textwidth}
\includegraphics[width=1\textwidth,angle=0]{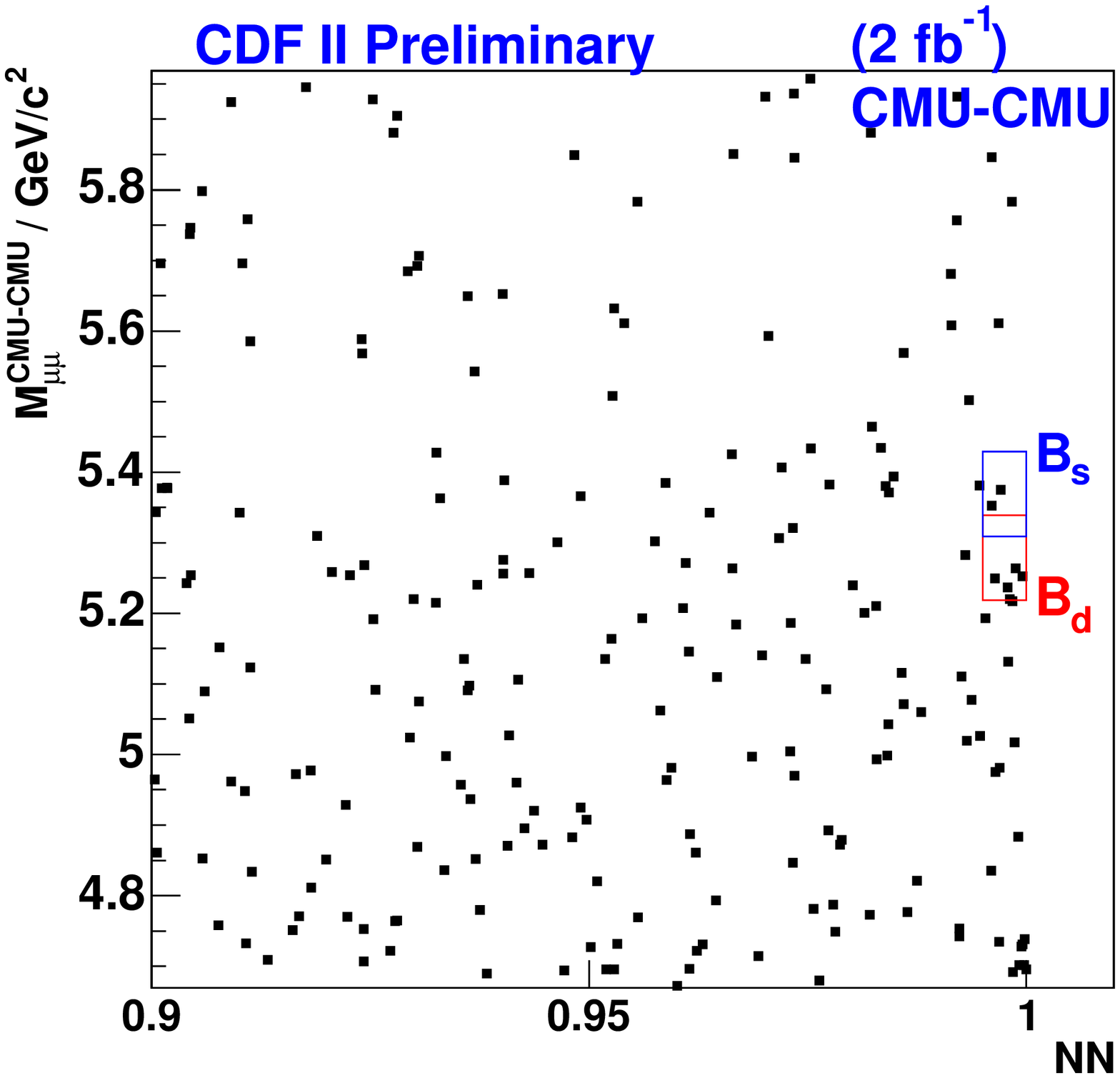}
\end{minipage}
\caption{The invariant mass distribution versus the NN output for the CMU-CMU channel\label{fig:12}.}
\end{center}
\end{figure}
\footnotetext[1]{The solid line is a graphical representation of the extracted 
yield, not a fit to the data.}
\subsection{$B^0_{s,d}\rightarrow \mu^+\mu^-$}
In case of the rare decays $B^0_{s,d}\rightarrow\mu^+\mu^-$ we use different neural network bins and mass bins for
the computation of the limits on the branching ratios. Table \ref{tab:1} shows the number of expected
and observed events of the two trigger scenarios CMU-CMU and CMU-CMX for different cuts on the network output NN 
and gives the result on the branching ratio for the combination of both scenarios with a neural network cut of $NN>0.80$.
In Figures \ref{fig:11} and \ref{fig:12} the invariant mass distribution vs. the neural network value is displayed. 
Using equation \ref{eqn2} we obtain the $90(95)\%$ C.L. limits 
\begin{eqnarray}
\mathcal{B}(B^0_s\rightarrow \mu^+\mu^-)<4.7\times
10^{-8} (5.8\times 10^{-8}) \nonumber \\
\mathcal{B}(B^0_d\rightarrow \mu^+\mu^-)<1.5\times 10^{-8} (1.8\times 10^{-8}) \nonumber
\end{eqnarray}
\section{Conclusion}
We present the latest results on the measurement of the
branching ratios of the rare decays $B\rightarrow \mu^+\mu^- h$
and give an upper limit on the branching ratio of the decays
$B^0_{s,d}\rightarrow\mu^+\mu^-$. The results for 
the branching ratios for the decay modes $B^+\rightarrow \mu^+\mu^- K^+$
and $B^0\rightarrow \mu^+\mu^- K^{*0}$ are in good agreement with the
results of the B factory experiments BABAR and BELLE \cite{BABAR,BELLE}. The limit on the branching
fraction $\mathcal{B}(B^0_s\rightarrow \mu^+\mu^- \phi)$/\\$
\mathcal{B}(B^0_s\rightarrow J/\psi \phi)$ is the most stringent to date.
The new results for the upper limit for the branching ratio of the decays
$B^0_{s,d}\rightarrow\mu^+\mu^-$ are currently the world's best limits
and can be used to reduce the allowed parameter space of a broad spectrum
of SUSY models \cite{SUSY01,SUSY02,SUSY03,SUSY04}.
%
\begin{table*}[t]
\begin{center}
\caption{Results of the $B\rightarrow \mu^+\mu^- h$ analysis.}
\label{tab:2}       
\vspace{0.25cm}
\begin{tabular}{llll}
\hline\noalign{\smallskip}
Mode & $B^+\rightarrow \mu^+\mu^- K^+$ & $B^0_d\rightarrow \mu^+\mu^- K^{*0}$ & $B^0_s\rightarrow \mu^+\mu^- \phi$ \\
\noalign{\smallskip}\hline\noalign{\smallskip}
$N_S$ & $90$ & $35$ & $11$ \\
$N_{BG}$ & $45.3\pm5.8$ & $16.5\pm3.6$ & $3.5\pm1.5$\\
Gaussian significance $(\sigma)$ & $4.5$ & $2.9$ & $2.4$ \\
Rel $\mathcal{B}\pm stat \pm sys \times 10^{-3}$ & $0.59\pm0.15\pm0.03$ & $0.62\pm0.23\pm0.07$ & $1.24\pm0.60\pm0.15$ \\
Abs $\mathcal{B}\pm stat \pm sys \times 10^{-6}$ & $0.60\pm0.15\pm0.04$ & $0.82\pm0.31\pm0.10$ & $1.16\pm0.56\pm0.42$ \\
Rel $\mathcal{B}$ 95\% CL limit $\times 10^{-3}$& - & - & $2.61$ \\
Rel $\mathcal{B}$ 90\% CL limit $\times 10^{-3}$& - & - & $2.30$ \\
\noalign{\smallskip}\hline
\end{tabular}
\end{center}
\end{table*}
\begin{table*}[t]
\begin{center}
\caption{Expected number of background (exp.) and number of observed (obs.) events in the CMU-CMU and CMU-CMX channels
for the $B^0_s$ mass window (5.310-5.430 GeV/c$^2$) and the $B^0_d$ mass window (5.219-5.339 GeV/c$^2$).}
\label{tab:1}       
\vspace{0.25cm}
\begin{tabular}{llll}
\hline\noalign{\smallskip}
 & Mode & $B^0_d\rightarrow \mu^+\mu^-$  & $B^0_s\rightarrow \mu^+\mu^-$ \\
\noalign{\smallskip}\hline\noalign{\smallskip}
CMU-CMU & N exp., $0.800<$NN$<0.950$ & $25.5\pm0.7$ & $23.5\pm0.7$   \\
CMU-CMU & N obs., $0.800<$NN$<0.950$ & 32  & 18 \\
CMU-CMU & N exp., $0.950<$NN$<0.995$ & $8.5\pm0.5$  & $7.7\pm0.5$   \\
CMU-CMU & N obs., $0.950<$NN$<0.995$ & 7 & 10 \\
CMU-CMU & N exp., $0.995<$NN & $2.4\pm0.2$  & $2.1\pm0.2$  \\
CMU-CMU & N obs., $0.995<$NN & 5 & 2 \\
\noalign{\smallskip}\hline\noalign{\smallskip}
CMU-CMX & N exp., $0.800<$NN$<0.950$ & $27.8\pm0.9$ & $26\pm0.7$\\
CMU-CMX & N obs., $0.800<$NN$<0.950$ & 28 & 26 \\
CMU-CMX & N exp., $0.950<$NN$<0.995$ & $10.8\pm0.5$   & $10.3\pm0.5$ \\
CMU-CMX & N obs., $0.950<$NN$<0.995$ &  6  & 11 \\
CMU-CMX & N exp., $0.995<$NN & $1.6\pm0.2$  & $1.6\pm0.2$  \\
CMU-CMX & N obs., $0.995<$NN & 1 & 1 \\
\noalign{\smallskip}\hline\noalign{\smallskip}
combined & Abs $\mathcal{B}$ 95\% CL limit $\times 10^{-8}$ & $1.8$ & $5.8$   \\
combined & Abs $\mathcal{B}$ 90\% CL limit $\times 10^{-8}$ & $1.5$ & $4.7$   \\
\noalign{\smallskip}\hline
\end{tabular}
\end{center}
\end{table*}
\section{Acknowledgments}
The author would like to thank the members of the 
CDF Collaboration who performed the analyses. \\ \\


\begin{thebibliography}{999}
\bibitem{ISO}
The B-candidate isolation is defined as $I=|\vec{p}_T^{\mu\mu}|/(\sum_i p_T^i + |\vec{p}^T|)$, where
the sum is over all tracks with $\sqrt{\Delta \eta^2 + \Delta \phi^2}\leq 1$; $\Delta \phi$ and $\Delta \eta$
are the azimuthal angle and pseudorapidity of track $i$ with respect to $\vec{p}^{\mu\mu}$.
\end{thebibliography}
\end{document}